\documentclass{elsart}

\usepackage{amssymb}
\usepackage{graphicx}

\begin{document}

\begin{frontmatter}
\title{STATISTICAL SCATTERING OF WAVES IN DISORDERED WAVEGUIDES:\\
Universal Properties}

\author
{P. A. Mello, M. Y\'epez}
\address
{Instituto de F\'isica, Universidad Nacional Aut\'onoma de M\'exico, 01000 M\'exico Distrito Federal, Mexico}

\author
{L. S. Froufe-P\'erez, J. J. S\'aenz}
\address{Departamento de F\'isica de la Materia Condensada and Instituto ``Nicol\'as Cabrera", Universidad Aut\'onoma de Madrid, E-28049 Madrid, Spain}

\begin{abstract}
The statistical theory of certain complex wave interference phenomena,
like the statistical fluctuations of transmission and reflection of waves,
is of considerable interest in many fields of physics.
In this article we shall be mainly interested in those situations where the complexity derives from
the quenched randomness of scattering potentials, as in the case of disordered conductors,
or, more in general, disordered waveguides.

In studies performed in such systems one has found remarkable
{\em statistical regularities},
in the sense that the probability distribution for various macroscopic quantities involves
a rather small number of relevant physical parameters, while the rest
of the microscopic details serves as mere ``scaffolding".
We shall review past work
in which this feature was captured following a maximum-entropy approach,
as well as later studies in which the existence of a limiting distribution,
in the sense of a generalized central-limit theorem,
has been actually demonstrated.
We then describe a microscopic potential model that was developed recently, which gives rise to a further generalization of the central-limit theorem and thus to a limiting macroscopic statistics.
\end{abstract}


\end{frontmatter}

\section{INTRODUCTION}
\label{intro}

Complex scattering of waves has captured the interest of physicists for a long time\cite{mello-kumar}.
For instance, the problem of coherent multiple scattering of waves, which has long been of great importance
in optics, has seen a revived interest in relation to the phenomenon of localization.

The present article fits in the general topic of
``Statistical theory of  complex wave-interference phenomena".
In particular, we shall study the statistical fluctuations of transmission and reflection of waves, which are
of considerable interest in mesoscopic physics.
Complexity in wave scattering may derive from:

\begin{itemize}

\item[(i)]
the chaotic nature of the underlying classical dynamics, as in
microwave cavities and quantum dots, or

\item[(ii)]
the randomness of the scattering potentials in a disordered medium, as
a disordered conductor, or a disordered waveguide carrying classical waves
(electromagnetic, elastic, etc.), on which we shall concentrate here.

\end{itemize}

Why do statistics on the results of a scattering process?
The point is that the interference pattern resulting from the
coherent multiple scattering of waves from the systems described above is so complex
(a small variation in some external parameter changes it completely)
that only a statistical treatment is meaningful.

We shall find a recurrent theme in our presentation:
the statistical regularity of the behavior, which involves a relatively small number of relevant physical
parameters, while the rest of details serves as mere
``scaffolding".
This feature was captured in the past following a Maximum-Entropy Approach, within the powerful,
non-perturbative, framework known as Random-Matrix Theory:
Shannon's information entropy is maximized, subject to the symmetries and constraints that are physically relevant
\cite{dmpk,mello-stone}.
Later, generalized Central-Limit Theorems (CLT) have been demonstrated
\cite{mello_shapiro,mello_tomsovic}.
Here we shall revisit past and recent efforts towards discovering
universal features in the statistical scattering of waves in disordered waveguides.

The paper is organized as follows. We first indicate the various physical
regimes to be encountered in the problem of disordered conductors. We then
mention how the statistical regularities in the problem have been captured
in the past within a Maximum-Entropy Approach: we briefly mention the transfer-matrix
method that was used and the Random-Matrix Theory model that was constructed,
giving rise to a diffusion equation in transfer-matrix space.
We then indicate a CLT that was
proved, thus showing that the Maximum-Entropy Approach captures the universal
features found in the CLT.
We then describe a microscopic potential model that was developed recently, which gives rise to a further generalization of the CLT and thus to a limiting macroscopic statistics. We then give our conclusions.

\section{THE MAXIMUM-ENTROPY APPROACH}
\label{system}

\begin{figure}[h]
\centering
\resizebox{10cm}{4cm}{\includegraphics{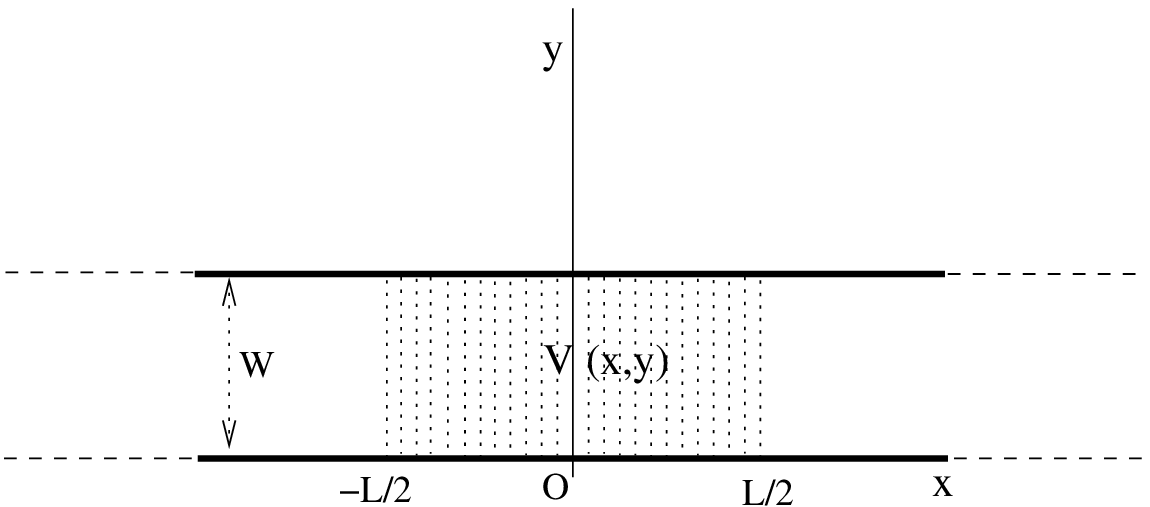}} (a)
\resizebox{10cm}{1cm}{\includegraphics{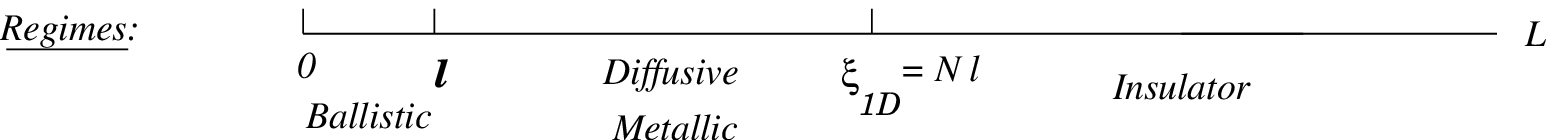}} (b)
\caption{
\footnotesize{A disordered waveguide and the various physical regimes.}
}
\label{q1d}
\end{figure}

The fundamental physical process occurring in the system under study,
the disordered waveguide shown schematically in Fig. \ref{q1d}(a), is coherent multiple scattering of waves.
The length of the disordered section of the waveguide is $L$, $W$ is its width
(the system being assumed two-dimensional) and $N$ is the total number of running modes supported by the waveguide; the mean-free-path (mfp) is denoted by $\ell$.
The system might as well be a disordered conductor with the same dimensions.
In the transport of waves through the system one encounters various physical regimes, indicated in Fig. \ref{q1d}(b):
i) the {\em ballistic} regime, for $0<L<\ell$;
ii) the {\em diffusive}, or {\em metallic} regime, for $\ell<L<N\ell$;
iii) the {\em insulating} regime, for  $N\ell<L$.

The aim is to calculate the transport properties of waves through this system.
One important quantity is, of course the conductance of the disordered quasi-one-dimensional (q-1d)
system, which is given by Landauer's formula:
\begin{eqnarray}
G&=&\frac{2e^2}{h}g \;, \\
g&=& {\rm tr}tt^{\dagger}
= \sum_{a,b =1}^N |t_{ab}|^2 \;,
\end{eqnarray}
where $a,b$ denote the $N$ transverse (running) modes: $a,b = 1, \cdots , N$.
Landauer's relation allows calculating the conductance --a transport quantity-- from the scattering properties of the system.
In microwave systems one can actually measure the individual transmission coefficients
$T_{ab}=|t_{ab}|^2$, as well as
$T= \sum_{a,b =1}^N |t_{ab}|^2$, experimentally.



The main approaches to the problem have been
\cite{mello-kumar,carlo97}:
i) Perturbation theory in the disordered potential;
ii) Supersymmetry methods, giving rise to a
non-linear sigma model and iii) Random-Matrix Theory (RMT) models of the
scattering matrix $S$, or the transfer matrix $M$ of the system. Here we shall
concentrate on these latter non-perturbative models.

One has observed remarkable {\em statistical regularities}, in the sense that
the probability distribution for various macroscopic quantities involves
a relatively small number of relevant physical parameters
(essentially the {\em mean free path} $\ell$).
Within a Random-Matrix Theory scheme,
this feature was captured in the past following a
{\em Maximum-Entropy Approach} \cite{dmpk},
which we now describe.

\begin{figure}[h]
\centering
\resizebox{10cm}{2.5cm}{\includegraphics{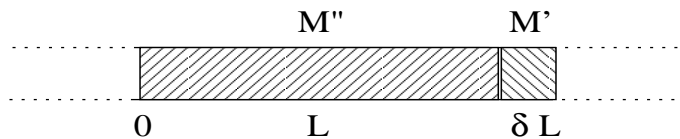}}
\caption{\footnotesize{A waveguide of length $L$ and a BB added to it.}}
\label{lbb}
\end{figure}

To a waveguide of length $L$ we assign the transfer matrix $M''$
(see Fig. \ref{lbb}), which has the property that acting on the wave amplitudes
$a^{(1)}$ on the left gives the amplitudes $a^{(2)}$ on the right:
\begin{equation}
M'' a^{(1)} = a^{(2)}.
\end{equation}
For every configuration of disorder we have one transfer matrix $M''$.
If we assign a probability density $p_L(M'')$ to our transfer matrices, what results is a {\em Random-Matrix Theory of transfer matrices}.

To the waveguide of length $L$ we now add a ``Building Block" (BB) of thickness
$\delta L$ which is much shorter than $L$, but {\em still contains many weak scatterers}
(the so called dense-weak-scattering limit (DWSL)), as shown schematically in
Fig. \ref{lbb}.

The transfer matrix $M'$ associated with the BB will be considered to be  statistically independent from $M''$;
we write its probability density as $p_{\delta L}(M')$.
The transfer matrices $M''$ and $M'$ are combined as
\begin{equation}
M = M' M''
\label{M=M'M''}
\end{equation}
to obtain the total transfer matrix of the combined system, while the probability density of the latter, $p_{L+\delta L}(M)$, is obtained from the individual ones by the ``convolution"
\begin{equation}
p_{L+\delta L}(M)
= \int p_L((M')^{-1}M) \; p_{\delta L}(M')\; d\mu (M')
\equiv p_L \star p_{\delta L} \; ,
\label{convolution}
\end{equation}
where $d\mu (M')$ is the invariant measure associated with the group of transfer matrices.
Eq. (\ref{convolution}) has the structure of the
{\em Smoluchowski equation in Brownian motion theory} \cite{chandra}.

Expecting the
{\em {results to be largely independent of the details of the BB}},
the distribution $p_{\delta L}(M')$ for the BB was modelled, in Ref. \cite{dmpk}, by
maximizing the {\em {Shannon entropy}}
\begin{equation}
\mathcal{S}[p] = - \int p(M') \ln p(M') dM' \; ,
\end{equation}
subject to the constraints
\begin{eqnarray}
&& \int p_{\delta L}(M') d\mu (M') =1 \;\;\;\;\; {\rm (normalization \;\; condition)}
\label{norm}
\\
&& \frac 1N \sum_{a,b=1}^N \langle|r_{ab}|^2 \rangle
=\frac{\delta L}{\ell} \;,
\label{constraint mfp}
\end{eqnarray}
$\ell$ being the elastic mean-free-path (mfp).
The only physical information conveyed by
$p_{\delta L}(M')$ is the mfp $\ell$ .

Following the procedure developed in the theory of Brownian motion, we can convert the integral equation (\ref{convolution}) into a differential equation, with the result \cite{dmpk,mello-stone}:
\begin{eqnarray}
&&
\frac{\partial w_s^{(\beta)}(\lambda)}{\partial s}
=
\frac{2}{\beta N +2 - \beta}
\nonumber \\
&&\;\;\;\;\;\times\sum_{a=1}^N \frac{\partial}{\partial \lambda_a}
\left[
\lambda_a(1+\lambda_a)J^{(\beta)}(\lambda)
\frac{\partial}{\partial \lambda_a}
\frac{w_s^{(\beta)}(\lambda)}{J^{(\beta)}(\lambda)}
\right] \;\; .
\label{diff eqn dmpk}
\end{eqnarray}
This is a {\em diffusion equation in transfer-matrix space}, known as the DMPK
equation (after Dorokhov \cite{Dorokhov} and Mello, Pereyra and Kumar
\cite{dmpk}), which governs the ``evolution"
with the length $L$ of the waveguide of the probability distribution
$w_s^{(\beta)}(\lambda)$ of our transfer matrices. This equation has been
written in terms of new variables as follows: i) $s = L/ \ell$ denotes the
length in units of the mfp; ii) $\lambda _a$ ($a=1,\cdots,N$) are ``radial"
variables in terms of which the conductance can be written as $g= \sum _{a=1}
^N 1/(1+\lambda _a)$; iii) we have used the Jacobian $J_ {\beta}(\lambda )
=\prod _{a<b}  \left| \lambda _a - \lambda _b \right| ^{\beta}$. Finally,
Eq. (\ref{diff eqn dmpk}) has to be solved with the initial condition $w_0
^{(\beta)}(\lambda ) = \delta (\lambda )$. The quantity $\beta$ ($=1,2,4$)
denotes the universality class of random-matrix theory \cite{dyson}.

\begin{figure}[h]
\centering
\resizebox{10cm}{8cm}{\includegraphics{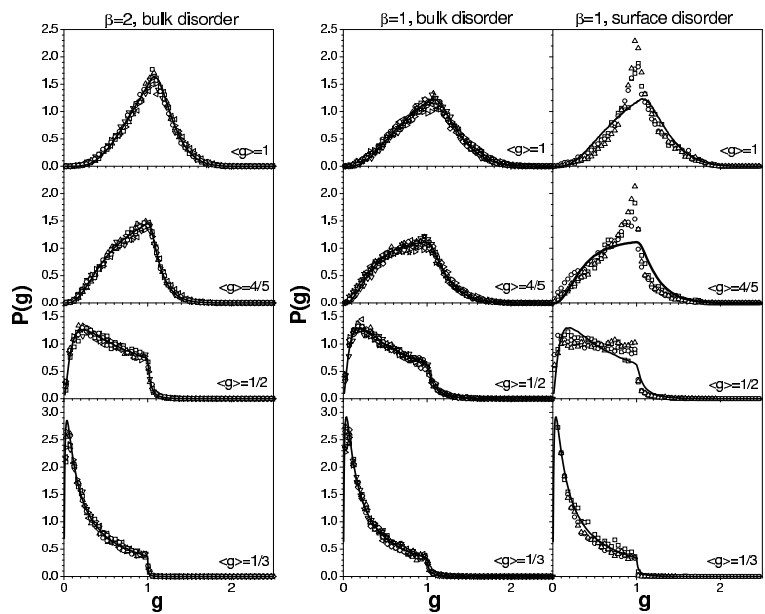}}
\caption{
\footnotesize{Conductance distribution from the DMPK equation and from
numerical simulations for bulk and surface disorder, as explained in the text
(after Ref. \cite{froufe_et_al_prl}).
}}
\label{dmpk vs anderson}
\end{figure}

The distribution $P(g)$ of the conductance is known to evolve from a Gaussian
(deep in the diffusive, metallic regime) to a log-normal distribution (deep in
the localized, insulating regime) \cite{carlo97}. Although $P(g)$ cannot be
easily obtained algebraically from the above expressions, various
approximations show that this behavior is well described by the DMPK equation
(\ref{diff eqn dmpk}). In the crossover regime, Ref. \cite{froufe_et_al_prl}
found the main statistical properties of $P(g)$ arising from the DMPK equation
using a Monte Carlo calculation. The results are shown in
Fig. \ref{dmpk vs anderson}, where $P(g)$ is plotted (solid lines) for both $\beta =1$ and 2 for
different values of $\langle g \rangle$.
The results of extensive q-1d
tight-binding-model calculations (symbols), carried out
for bulk disorder, are also shown.
We see that the agreement between the two types of results is excellent.
In contrast, the results for surface disorder and $\beta =1$, also shown in the figure
for comparison, are not described properly by the DMPK equation.

\section{CENTRAL-LIMIT THEOREMS}
\label{clt}

It was shown later \cite{mello_shapiro} that a limiting distribution
for $w_s(\lambda)$ arises
when the individual, microscopic, scattering units are combined in the so
called {\em dense-weak-scattering limit} (DWSL) and within a particular class
of models. The DWSL corresponds to a large density of weak scatterers, with a
fixed mfp $\ell$. In the particular model that was studied, the individual
scattering units were defined through their transfer matrices $M_i$ and an
``isotropic" distribution of their phases was assumed. The limiting
distribution that arises depends only on $\ell$ and is insensitive to other
details of the microscopic distribution: it thus constitutes a {\em generalized
central-limit theorem} (CLT). The result turns out to be identical to the DMPK
equation (\ref{diff eqn dmpk}) found in the maximum-entropy model described
above. We can thus say that {\em the maximum-entropy model selects the limiting
distribution}, in the sense of the DWSL, within a class of models
for the transfer matrices of the individual scattering units.

A class of limiting distributions wider than that of Ref. \cite{mello_shapiro} was studied by one of the present authors (PAM) and S. Tomsovic in
Ref. \cite{mello_tomsovic}, in which
the isotropy assumption of Ref. \cite{mello_shapiro} was relaxed to a large extent.
In Ref. \cite{mello_tomsovic} the DWSL plays again an essential role and the result is a more general CLT than that of Ref. \cite{mello_shapiro}.
The evolution with $L$ is described by a generalized diffusion equation, in which the diffusion coefficients are
the inverse mfp's for the various scattering processes that may occur in the
problem.
When the various mfp's can be represented by a single one, one encounters the
DMPK equation that was described above.
Thus the model of Ref. \cite{mello_tomsovic} appears as a possible candidate to study the influence of the specific scattering properties of the various modes, which seem to be relevant, for instance, for the problem of waveguides with surface disorder, where DMPK does not give a proper description.

\section{A POTENTIAL MODEL FOR THE EVOLUTION OF EXPECTATION VALUES.
A CENTRAL-LIMIT THEOREM}
\label{evol<>}

We first present a general way of expressing the expectation value of an observable when we add a BB to an already existing waveguide of length $L$, as shown in Fig. \ref{lbb}.
The transfer matrix of the two pieces is combined as in Eq. (\ref{M=M'M''}), which can also be written as
\begin{equation}
M=M'' + \delta M
=M'' + \varepsilon M'' \; ,
\end{equation}
where we have expressed the transfer matrix of the BB as
\begin{equation}
M' = I + \varepsilon.
\label{M'=I+vareps}
\end{equation}
Consider now a function $F(M)$ of the transfer matrix $M$, whose statistical properties we want to study: it might be, for instance, the conductance $G$ studied earlier, the transmission coefficient $T_{ab}$, or any other quantity of physical interest.
Its average for the enlarged piece can be written in terms of that for the original one as
\begin{eqnarray}
&&\left\langle F(M)\right\rangle _{L+\delta L}
=\left\langle F(M)\right\rangle _L
+\left\langle \varepsilon^{\cdot}\right\rangle _{\footnotesize L, \delta L}
\left\langle M^{\cdot}
\frac{\partial F(M)}{\partial M^{\cdot}} \right\rangle _L
\nonumber \\ \nonumber \\
&&\;\;\;\;+\;\frac 1{2!}
\left\langle
\varepsilon^{\cdot} \varepsilon ^{\cdot\cdot}
\right\rangle_{L, \delta L}
\left\langle
M^{\cdot} M^{\cdot\cdot}
\frac{\partial ^2F(M)} {\partial M^{\cdot}M^{\cdot\cdot}}
\right\rangle _L
+\cdots .
\label{<>L+dL}
\end{eqnarray}
To simplify the notation, we have indicated symbolically with dots
the relevant summations over channel and block indices
(see Ref. \cite{mello_tomsovic} for more details).

\begin{figure}[h]
\centering
\resizebox{15cm}{1cm}{\includegraphics{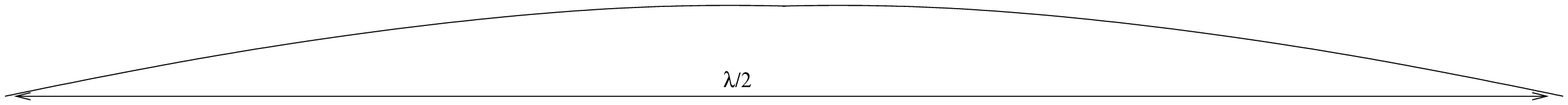}}
\resizebox{7cm}{3cm}{\includegraphics{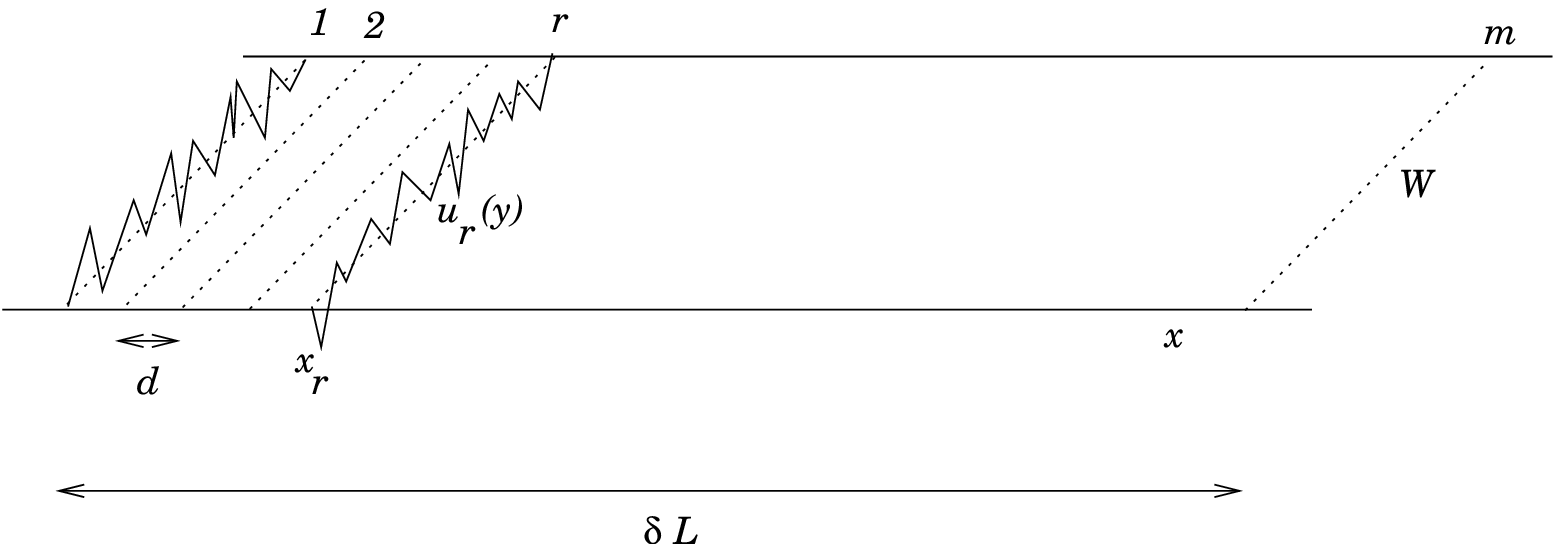}}
\caption{
\footnotesize{Construction of the BB using $\delta$-potential slices in the regime defined by the inequalities (\ref{regime}).}
}
\label{random_slices}
\end{figure}

It is clear that we now need an expression for the various moments of the quantities $\varepsilon$ associated with the BB.
Recently \cite{luis_miztli_pier_juanjo}, such an expression has been obtained
from a potential model, which we now outline.
We construct the BB as a sequence of $m\gg 1$ random $\delta$-potential slices,
such that
\begin{equation}
d \ll \delta L \ll \{ \lambda , \ell \} ,
\label{regime}
\end{equation}
as shown schematically in Fig. \ref{random_slices}.

The $r$-th $\delta$-slice potential, defined as
\begin{equation}
U_r(x,y) = u_r(y)\delta (x-x_r) ,
\label{r delta slice (xy)}
\end{equation}
has matrix elements with respect to channels given by
\begin{equation}
\left[ U_r(x) \right]_{ab}
= {(u_r)}_{ab}\delta (x-x_r),
\end{equation}
in terms of which we specify the statistical model.
The $m$ potentials $u_r$, $r=1,\cdots,m$, are assumed to be
{\em statistically independent},
{\em identically distributed, with zero average} and, for simplicity,
{\em zero odd moments}, so that
\begin{eqnarray}
&&\langle  (u_r)_{ab} \rangle
= 0
\\
&&\Big\langle  (u_{r_1})_{a_1 b_1} (u_{r_2})_{a_2 b_2}\Big\rangle
= \kappa _2(a_1 b_1, a_2 b_2) \delta _{r_1 r_2}
\\
&&\Big\langle  (u_{r_1})_{a_1 b_1} (u_{r_2})_{a_2 b_2}
(u_{r_3})_{a_3 b_3} (u_{r_4})_{a_4 b_4}
\Big\rangle
=
\kappa _2 (a_1 b_1, a_2 b_2)\kappa _2 (a_3 b_3, a_4 b_4)
\delta _{r_1 r_2}\delta _{r_3 r_4}
\nonumber \\
&&\hspace{1cm}+\kappa _2 (a_1 b_1, a_3 b_3)\kappa _2 (a_2 b_2, a_4 b_4)
\delta _{r_1 r_3}\delta _{r_2 r_4}
\nonumber \\
&&\hspace{2cm}+\kappa _2 (a_1 b_1, a_4 b_4)\kappa _2 (a_2 b_2, a_3 b_3)
\delta _{r_1 r_4}\delta _{r_2 r_3}
\nonumber \\
&&\hspace{3cm}
+\kappa _4 (a_1 b_1, a_2 b_2, a_3 b_3, a_4 b_4)\delta _{r_1 r_2 r_3 r_4}
\\
&&\hspace{4cm} \cdots ,
\nonumber
\end{eqnarray}
where
$\kappa _2(a_1 b_1, a_2 b_2), \kappa _4 (a_1 b_1, a_2 b_2, a_3 b_3, a_4 b_4)$,
etc., denote the second, fourth, etc., cumulants of
$(u_r)_{ab}$.
From these expressions we can calculate, in the DWSL, the various moments of $\varepsilon$ needed in Eq. (\ref{<>L+dL}).
One finds that the first moment vanishes, the second moment behaves linearly
with $\delta L$ and higher moments behave as higher powers thereof.
Also, the very important result emerges that {\em the dependence on the cumulants of the potential higher than the second drops out in the DWSL}.
The {\em diffusion coefficients}
$D_{ab,cd}^{jk,lm}$, or inverse mfp's $\ell_{ab}$, defined as
\begin{equation}
\left\langle
\varepsilon _{ab}^{jk}  \varepsilon _{cd}^{lm}
\right\rangle_{L, \delta L}
= 2D_{ab,cd}^{jk,lm}(k,L) \; \delta L + \cdots ,
\label{general_diff_coeff}
\end{equation}
depend only upon the second cumulants of the potential.
They are energy dependent and also length dependent.

Finally, we take the first term on the r.h.s. of Eq. (\ref{<>L+dL}) to the l.h.s.,
divide both sides by $\delta L$ and take the limit $\delta L \to 0$.
The result is the Fokker-Planck equation:
\begin{equation}
\frac{\partial \left\langle F(M)\right\rangle_L}{\partial L}
=D_{ab,cd}^{jk,lm}
\left\langle M_{be}^{kn}M_{df}^{mp}\frac{\partial ^2F}
{\partial M_{ae}^{jn}\partial M_{cf}^{lp}}\right\rangle_L  .
\label{general_diff_eqn}
\end{equation}
The fact that cumulants of the potential higher than the second
are irrelevant in the end signals the existence of a {\em generalized CLT}:
once the mfp's are specified, the limiting equation (\ref{general_diff_eqn})
is {\em universal}, i.e., independent of other details of the microscopic statistics.

One of the main difficulties in solving Eq. (\ref{general_diff_eqn}), both analytically and numerically, is that it involves averages of different quantities on the l.h.s. and on the r.h.s.
So far, that equation has been solved analytically for the one-open-channel case ($N=1$) and a restricted number of ``observables"
only: the results thus obtained are in excellent
agreement with microscopic calculations  \cite{luis_miztli_pier_juanjo}.
Numerically, we have found no ``direct" way of solving
Eq. (\ref{general_diff_eqn}).
Recently, a numerical algorithm which was called ``random walk in transfer-matrix space" has been implemented
\cite{luis_miztli_pier_juanjo,luis_thesis},
in which a BB is constructed with the property given in
Eq. (\ref{general_diff_coeff}), and then combined with successive BB's to construct a waveguide of finite length $L$.
The results have been compared with those arising from microscopic calculations, in which the entities that are combined are, literally, individual potential slices. For situations in which we have bulk disorder, the comparison is excellent, even for quantities which are not described properly by DMPK.
For surface disorder, preliminary results indicate a reasonable agreement for the observables examined so far.

\section{CONCLUSIONS}
\label{concl}

In this paper we have first revisited some earlier results which seem to indicate that a maximum-entropy approach to the problem of transport in disordered waveguides works well when there is a central-limit theorem (CLT) ``behind the scenes".

We have briefly described a rather general CLT that was obtained in the past, in Ref. \cite{mello_tomsovic}.

Recently, we have shown that a CLT arises in a model consisting of a random
distribution of $\delta$-potential slices. The parameters on which the result
depends are the mfp's $\ell_{ab}$ --which depend on the variances of the
potential-slices matrix elements-- and the correlation coefficient of these
matrix elements. Other details of the potential distribution are ``washed out"
in the dense-weak-scattering limit (DWSL). The result is expressed in terms of
a generalized diffusion equation for the evolution with length of expectation
values of physical observables.

Numerical results based on the ``random walk in transfer-matrix space" method
indicate an excellent agreement with microscopic calculations with bulk disorder, and a reasonable one for the problem with surface disorder.

More effort is needed towards an analytical, as well as a numerical, treatment
of the diffusion equation.

P.A.M. and M.Y. acknowledge financial support by Conacyt, Mexico, the former through grant No. 42655.
L.S.F. and J.J.S. have been supported by the Spanish MCyT (Ref. No.
BFM2003-01167) and the EU Integrated Project ``Molecular Imaging'' (EU contract
LSHG-CT-2003-503259).

\end{document}